\documentclass[conference]{IEEEtran}
\IEEEoverridecommandlockouts
\usepackage{cite}
\usepackage{svg}
\usepackage{amsmath,amssymb,amsfonts}
\usepackage{algorithmic}
\usepackage{graphicx}
\usepackage{textcomp}
\usepackage{tabularx}
\usepackage[margin=1in]{geometry}
\usepackage{booktabs}
\usepackage{multirow}
\usepackage{xcolor}
\usepackage{longtable}
\usepackage{array}
\usepackage{multirow}
\def\BibTeX{{\rm B\kern-.05em{\sc i\kern-.025em b}\kern-.08em
    T\kern-.1667em\lower.7ex\hbox{E}\kern-.125emX}}

\usepackage{array}
\usepackage{longtable}
\usepackage{geometry}
\usepackage{lscape}
\usepackage{graphicx}
\usepackage{url}
\usepackage{doi}

\newcolumntype{C}[1]{>{\centering\arraybackslash}m{#1}}    
\begin{document}
\pagestyle{plain}

\title{Risk Assessment and Threat Modeling for safe autonomous driving technology\\

\thanks{Identify applicable funding agency here. If none, delete this.}
}

\author{
\IEEEauthorblockN{Ian Alexis Wong Paz}
\IEEEauthorblockA{\textit{Department of Computer Science} \\
\textit{Florida Gulf Coast University} \\
Fort Myers, Florida \\
iawongpaz7677@eagle.fgcu.edu}
\and
\IEEEauthorblockN{Anuvinda Balan}
\IEEEauthorblockA{\textit{Department of Computer Science} \\
\textit{Florida Gulf Coast University} \\
Fort Myers, Florida \\
abalan5892@eagle.fgcu.edu}
\and
\IEEEauthorblockN{Sebastian Campos}
\IEEEauthorblockA{\textit{Department of Computer Science} \\
\textit{Florida Gulf Coast University} \\
Fort Myers, Florida \\
srcampos8611@eagle.fgcu.edu}
\and
\IEEEauthorblockN{Ehud Orenstain}
\IEEEauthorblockA{\textit{Department of Computer Science} \\
\textit{Florida Gulf Coast University} \\
Fort Myers, Florida \\
eorenstain5901@eagle.fgcu.edu}
\and
\IEEEauthorblockN{Sudip Dhakal}
\IEEEauthorblockA{\textit{Department of Computer Science} \\
\textit{Florida Gulf Coast University} \\
Fort Myers, Florida \\
sdhakal@fgcu.edu}
}

\maketitle
\begin{table*}
\centering
\caption{Elements of the Autonomous Vehicle Architecture}
\begin{tabularx}{\textwidth}{|l|l|l|X|}
\hline
\textbf{Sub-sections} & \textbf{Type of Element} & \textbf{System Assets} & \textbf{Attacks Through System Assets and Entry Points} \\
\hline
\multirow{6}{*}{Device \& Peripheral} 
& Software & Sensors & Influence of spoofing,  harvesting, internal attack, malicious code, outdated software \\
& Wireless Communication & Modem  & Denial of Service (DoS), Man-in-the-middle, induce misleading data from spoofing\\
& Radar & Sensors & Spoofing attacks, outdated software, malicious code, malware of a smartphone  \\
& GPS & Software & Spoofing fake signals, data injection, exploit outdated software \\
& Sensor Data Storage & Data Storage & Spoofing, malware of a smartphone, Outdated Software attacks \\
\hline
\multirow{10}{*}{Connected Vehicle} 
& Cloud Services & Data stored in the cloud & Malicious software used to find the vulnerabilities in the cloud access controls  \\
& Decision Maker & Machine Learning And AI & Malicious injection code to the vulnerabilities of connected software components \\
& CAN Bus & ECU & Overloading the network by an DoS attack to the ECU \\
& Control ECU & Software & Spoofing the ECU by sending compromised data\\
& Sensor Fusion & Memory & Malicious updates through sensors \\
& Perception AI & Machine Learning Model & Overloading the hardware used by using spoofing on the AI \\
& Path Planner & Software & Exploiting the weakness in the code by introducing bugs and malicious code\\
\hline
\end{tabularx}
\end{table*}
\begin{abstract}
This research paper delves into the field of autonomous vehicle technology, examining the vulnerabilities inherent in each component of these transformative vehicles. Autonomous vehicles (AVs) are revolutionizing transportation by seamlessly integrating advanced functionalities such as sensing, perception, planning, decision-making, and control. However, their reliance on interconnected systems and external communication interfaces renders them susceptible to cybersecurity threats.

This research endeavors to develop a comprehensive threat model for AV systems, employing OWASP Threat Dragon and the STRIDE framework. This model categorizes threats into Spoofing, Tampering, Repudiation, Information Disclosure, Denial of Service (DoS), and Elevation of Privilege.

A systematic risk assessment is conducted to evaluate vulnerabilities across various AV components, including perception modules, planning systems, control units, and communication interfaces. 

\end{abstract}


\section{Introduction}
This comprehensive paper presents a detailed analysis of autonomous vehicle technology and its associated security considerations. The paper commences by tracing the progressive evolution of autonomous vehicles, from rudimentary driver assistance features to sophisticated sensor suites. It underscores the potential of autonomous vehicles to mitigate human error, enhance traffic efficiency, and expand mobility for individuals without access.

The paper emphasizes the imperative of adopting a systematic security approach that aligns with technological advancements.

The heart of autonomous vehicle systems lies in the sensing and perception modules. These modules employ high-resolution cameras, LiDAR including millimeter wave variants, multiple radar systems, GPS, and vehicle-to-everything communications. These components collaborate to construct a dynamic 3D perception of the road environment. Data fusion algorithms harmonize visual images, point cloud signals, and network position updates, enabling artificial intelligence models to discern and categorize pedestrians, vehicles, traffic signs, lane markings, and environmental conditions such as rain, fog, or low lighting[1].

Building upon the perception capabilities, the planning and decision modules transform the understanding of the scene into safe and lawful motion. Planners assess potential routes around stationary or moving objects and compute trajectories that adhere to vehicle dynamics and traffic regulations[22]. Subsequently, decision logic selects maneuvers such as lane changes, speed adjustments, or emergency braking by considering the anticipated behavior of other road users and regulatory constraints. The control subsystem then dispatches precise real-time commands to steering, throttle, and braking units, executing the selected path.

Furthermore, the paper addresses adaptive control loops and fallback strategies that ensure stability and comfort under unforeseen circumstances.

In the context of advancing technological advancements, we present a structured risk assessment and threat model specifically tailored for autonomous vehicles. This comprehensive analysis systematically evaluates various potential risks that could compromise the safety of autonomous vehicles. These risks encompass environmental factors such as glare and precipitation, sensor attacks including laser blinding and fake object projection, and network exploits such as fraudulent vehicle-to-everything messages and software vulnerabilities.

Each identified risk is meticulously assessed based on its likelihood and potential impact. Consequently, we propose a comprehensive suite of countermeasures to mitigate these risks. These countermeasures include robust AI training, multiple sensor redundancy, secure communication protocols, trusted boot for controllers, and continuous network monitoring.

Furthermore, we provide a detailed structure of the paper and contextualize our contributions within the existing research landscape. Section two reviews related work in the field of perception and vehicle security. Section three presents our functional decomposition of the sensing perception planning decision-making, and control processes. Section four outlines the threat taxonomy and risk matrix. Section five proposes mitigation strategies and describes our threat modeling methodology. Finally, Section six concludes with recommendations for future investigations and best practices for ensuring the safe and reliable deployment of autonomous mobility. By seamlessly integrating technical innovation with a holistic security perspective, we pave the way for resilient and trustworthy autonomous mobility, paving the way for its widespread adoption by the public.

\begin{figure}[htbp]
    \centering
    \includegraphics[width=0.9\linewidth]{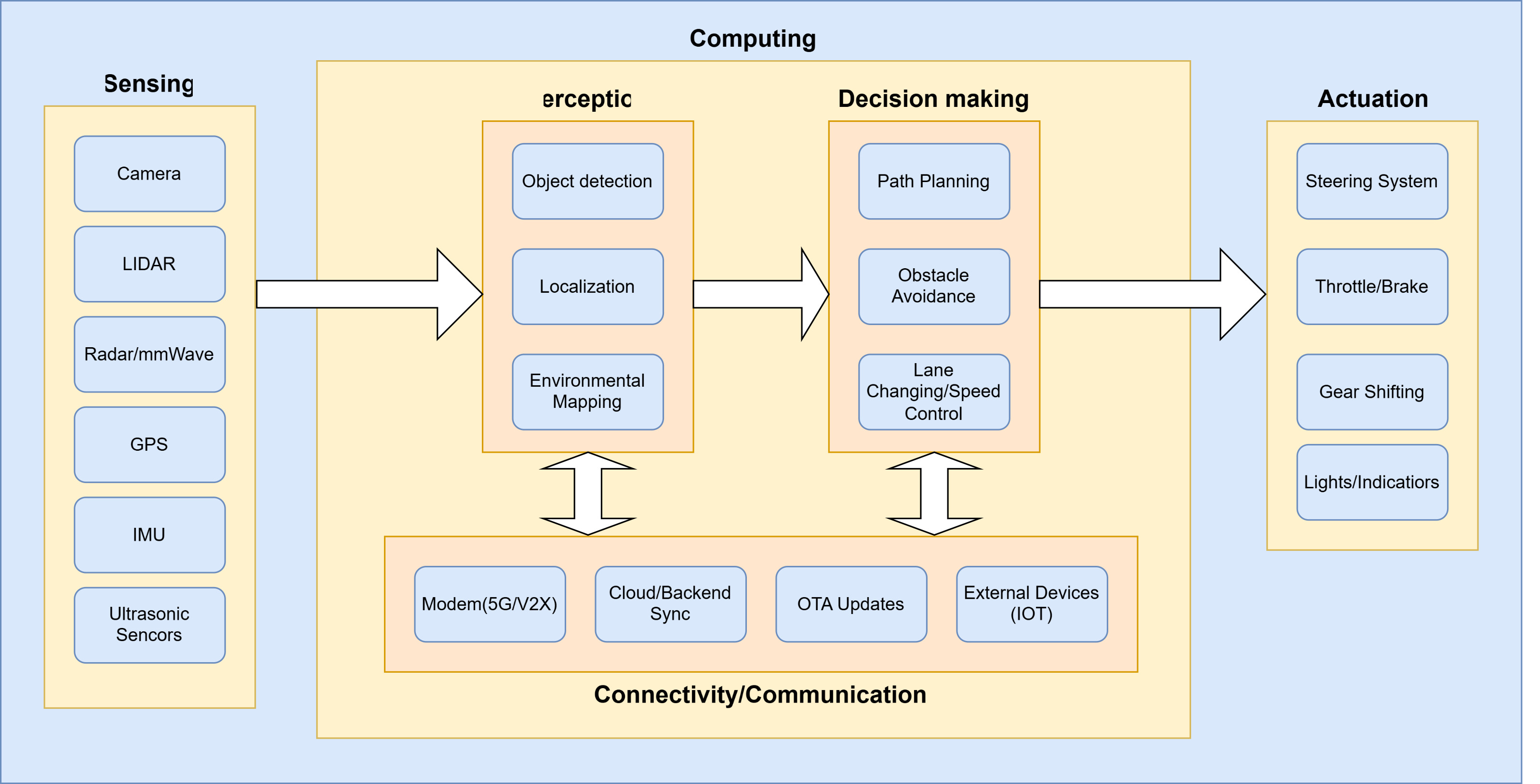}
    \caption{Autonomous vehicles components}
    \label{fig:fig1}
\end{figure}

\section{Risk Associated}

Despite rapid advancements in sensing, perception, planning, and control, autonomous vehicles remain vulnerable to a diverse range of threats that can compromise steering, braking, and overall safety.

Spoofing Attacks

Spoofing attacks impersonate trusted components or signals to deceive perception and decision modules. Examples include spoofing a sensor’s identity, GPS spoofing, spoofing the autonomous-driving ECU identity, fake V2V messages, man-in-the-middle on V2X, voice-command exploits, and false firmware-update notifications. An adversary may broadcast counterfeit GPS data to divert a vehicle onto hazardous roads or manipulate vehicle-to-everything messages to simulate nonexistent traffic. Sensor-identity spoofing can cause the system to accept erroneous camera or radar data, while forged firmware updates install malicious software. Voice-command exploits trigger unintended maneuvers by mimicking authorized speech. Defenses necessitate cryptographic authentication of all signals, certificate-based message verification, and continuous anomaly detection.

Tampering Attacks

Tampering attacks disrupt data streams or the physical environment to subvert normal operation. Examples include tampering with message content, tampering with the environment, smartphone spoofing, machine-learning tampering, camera image tampering, adversarial visual input, insecure API exposure, roadside-unit compromise, and cloud-to-AV sync attacks [2,3,4,10,11,12,13,19]. An attacker may inject malicious content onto internal buses to alter planned trajectories or disable alerts during critical maneuvers [11]. Physical tampering—such as deploying fog machines, light-rain sprays, or artificial snow—obscures obstacles and distorts perception [2,3,4]. Compromised roadside units or exposed APIs enable remote code injection or data exfiltration [13,19]. Robust end-to-end integrity checks, stringent input validation, real-time model monitoring, and hardened interfaces are essential countermeasures [11,12,19].

Denial of Service Threats

Denial-of-service (DoS) attacks overwhelm or disable sensors, networks, and processing elements, disrupting vital data flows. Examples include flooding the inner vehicle network, DDoS modem attacks, LiDAR overload, software denial of service, LiDAR sensor blinding, radar jamming, DoS on the CAN bus, and physical destruction of sensors [6,7,10,14]. Flooding the internal communication bus prevents safety alerts from reaching controllers [10], while modem-based DoS isolates the vehicle from over-the-air updates [10]. LiDAR overload or blinding negates depth perception [7], and radar jamming masks Doppler returns [6]. Targeted electromagnetic interference or outright destruction of sensors can eliminate autonomy-critical inputs [14]. Resilience techniques include redundant communication channels, sensor health-check routines, and graceful degradation strategies [7,14].

Information-Disclosure Attacks

Information-disclosure attacks involve the unauthorized acquisition or dissemination of sensitive data from sensors, networks, or cloud platforms. These attacks encompass various tactics, such as stealing sensor information, exploiting prior knowledge vulnerabilities, causing sensor-fusion conflicts, breaching cloud data, transmitting unencrypted V2X communications, and violating privacy through telemetry data. Intercepting raw sensor feeds can reveal detailed maps of passenger locations and travel patterns. Additionally, compromised supply-chain elements or insider actors may leak proprietary control logic or encryption keys. Unencrypted telemetry streams enable adversaries to reconstruct routes and behavior profiles. Effective mitigations include implementing robust encryption for both data at rest and in transit, enforcing strict access controls, and maintaining continuous audit logging.

Elevation of Privilege Threats

Elevation-of-privilege attacks grant adversaries elevated system privileges to manipulate vehicle behavior. These attacks encompass various tactics, including remote code injection via over-the-air (OTA) updates, braking system hijacking, steering manipulation, physical ECU tampering, and insider threats within the manufacturing process. Remote code injection can seize control of braking or steering subsystems without physical proximity. Physical tampering with ECUs may bypass safety interlocks or install persistent malicious firmware. Insider threats during manufacturing can implant hidden backdoors that evade authentication checks. Defenses against these threats include secure-boot processes, hardware root-of-trust anchors, multi-factor authentication for maintenance interfaces, and rigorous supply-chain verification.

To effectively counter this extensive threat spectrum, a multi-layered defense strategy is essential. Redundant sensor fusion with overlapping fields of view ensures sustained perception even if one sensor channel is compromised. Training AI models on a broad range of real-world and simulated spoofing scenarios enhances resilience against adversarial inputs. Securing all communication channels with robust authentication and encryption prevents unauthorized message injection. Verifying firmware integrity before execution and enforcing trusted-boot processes prevent remote code injection via over-the-air updates. Continuous system monitoring combined with safe fallback procedures—such as gradually decelerating the vehicle or transferring control to a prepared human operator—preserves safety even under active attack.

\section{Risk Analysis}
During our research, we conducted a systematic risk assessment by compiling all identified threats and vulnerabilities into a comprehensive table using the DREAD scoring methodology. This approach allowed us to evaluate each risk across five key dimensions: Damage potential, Reproducibility, Exploitability, Affected Users, and Discoverability. By assigning numerical scores to each dimension, we were able to quantitatively rank the severity and potential impact of each threat on the autonomous vehicle system. The resulting table, presented in the following chapter, provides an in-depth explanation of each threat, describing not only its technical nature but also the specific attack vectors that adversaries might exploit.The table also highlights mitigation strategies for each risk, providing actionable solutions to enhance vehicle security. This evaluation helps prioritize defenses and inform future system improvements.

\onecolumn

\begin{longtable}{|>{\centering\arraybackslash}p{0.3cm}|>{\centering\arraybackslash}p{2.8cm}|>{\centering\arraybackslash}p{1.2cm}|>{\centering\arraybackslash}p{2cm}|>{\centering\arraybackslash}p{2cm}|>{\centering\arraybackslash}p{1.5cm}|>{\centering\arraybackslash}p{2cm}|>{\centering\arraybackslash}p{1.4cm}|}
\caption{DREAD based scoring details} \label{tab:dread-scoring-details} \\
\hline
\textbf{ID} & \textbf{Description} & \textbf{Damage} & \textbf{Reproducibility} & \textbf{Exploitability} & \textbf{Affected Users} & \textbf{Discoverability} & \textbf{DREAD score} \\
\hline
\endfirsthead

\multicolumn{8}{c}{\tablename\ \thetable\ -- \textit{Continued from previous page}} \\
\hline
\textbf{ID} & \textbf{Description} & \textbf{Damage} & \textbf{Reproducibility} & \textbf{Exploitability} & \textbf{Affected Users} & \textbf{Discoverability} & \textbf{DREAD score} \\
\hline
\endhead

\hline \multicolumn{8}{r}{\textit{Continued on next page}} \\
\endfoot

\hline
\endlastfoot 
1 & Spoofing a sensors identity & 4 & 4 & 3 & 1 & 3 & 3 \\ \hline
2 & Spoofing the Autonomous driving ECU identity & 4 & 4 & 3 & 1 & 3 & 3 \\ \hline
3 & Tampering with the message content & 4 & 4 & 3 & 1 & 3 & 3 \\ \hline
4 & Stealing sensor information & 4 & 4 & 3 & 1 & 3 & 3 \\ \hline
5 & Flooding the inner vehicle network & 4 & 4 & 3 & 1 & 3 & 3 \\ \hline
6 & Destroying the sensors & 2 & 4 & 3 & 1 & 4 & 2.8 \\ \hline
7 & Tampering with the enviroment & 4 & 4 & 1 & 1 & 4 & 2.8 \\ \hline
8 & DDOS Modem & 4 & 2 & 3 & 3 & 1 & 2.6 \\ \hline
9 & Overload LiDAR & 4 & 2 & 3 & 2 & 1 & 2.5 \\ \hline
10 & Prior Knowledge & 4 & 3 & 3 & 3 & 3 & 3.2 \\ \hline
11 & Machine learning tampering & 4 & 2 & 3 & 2 & 1 & 2.4 \\ \hline
12 & Smartphone spoofing & 4 & 2 & 3 & 3 & 2 & 2.8 \\ \hline
13 & Spoofing Snow & 4 & 2 & 3 & 2 & 2 & 2.5 \\ \hline
14 & Spoofing Rain droplets & 4 & 2 & 3 & 2 & 2 & 2.6 \\ \hline
15 & Fog spoofing & 4 & 2 & 3 & 2 & 2 & 2.6 \\ \hline
16 & Tampered sensors data & 4 & 4 & 3 & 1 & 2 & 2.8 \\ \hline
17 & Buffer overflow attack on the Modem & 4 & 4 & 3 & 3 & 2 & 2.6 \\ \hline
18 & Software Denial of Service & 4 & 4 & 3 & 1 & 3 & 3 \\ \hline
19 & GPS Spoofing & 4 & 3 & 4 & 4 & 4 & 3.8 \\ \hline
20 & LiDAR Sensor Blinding & 3 & 4 & 3 & 4 & 3 & 3.4 \\ \hline
21 & Camera Image Tampering & 4 & 3 & 4 & 4 & 2 & 3.4 \\ \hline
22 & Radar Jamming & 4 & 3 & 2 & 4 & 4 & 3.4 \\ \hline
23 & Sensor Fusion Conflict & 3 & 4 & 3 & 4 & 4 & 3.6 \\ \hline
24 & Adversarial Visual Input & 3 & 4 & 4 & 4 & 4 & 3.8 \\ \hline
25 & Remote Code Injection via OTA & 4 & 4 & 2 & 3 & 4 & 3.4 \\ \hline
26 & Man-in-the-Middle (MITM) on V2X & 4 & 3 & 4 & 4 & 2 & 3.4 \\ \hline
27 & Denial of Service on CAN Bus & 4 & 4 & 3 & 4 & 4 & 3.8 \\ \hline
28 & Decision AI Model Poisoning & 4 & 4 & 3 & 4 & 4 & 3.8 \\ \hline
29 & Path Planning Misguidance & 4 & 4 & 2 & 4 & 2 & 3.2 \\ \hline
30 & Braking System Hijack & 4 & 4 & 3 & 4 & 2 & 3.4 \\ \hline
31 & Steering Manipulation & 3 & 4 & 4 & 4 & 2 & 3.4 \\ \hline
32 & Fake V2V Message & 4 & 3 & 3 & 4 & 2 & 3.2 \\ \hline
33 & Cloud Data Breach & 4 & 3 & 4 & 4 & 4 & 3.8 \\ \hline
34 & Insecure API Exposure & 3 & 4 & 4 & 4 & 3 & 3.6 \\ \hline
35 & Unencrypted V2X Transmission & 4 & 3 & 4 & 4 & 3 & 3.6 \\ \hline
36 & Roadside Unit Compromise & 4 & 4 & 4 & 4 & 2 & 3.6 \\ \hline
37 & Physical ECU Tampering & 4 & 4 & 4 & 3 & 3 & 3.6 \\ \hline
38 & Insider Threat in Manufacturing & 4 & 4 & 3 & 4 & 3 & 3.6 \\ \hline
39 & Voice Command Exploit & 3 & 4 & 2 & 3 & 2 & 2.8 \\ \hline
40 & Cloud-to-AV Sync Attack & 3 & 4 & 4 & 3 & 3 & 3.4 \\ \hline
41 & Privacy Violation via Telemetry Data & 4 & 3 & 3 & 4 & 4 & 3.6 \\ \hline
42 & False Firmware Update Notification & 4 & 4 & 4 & 4 & 3 & 3.8 \\ \hline

\end{longtable}

\clearpage
\onecolumn

\section{Threat Modeling}

\begin{longtable}{|C{0.3cm}|C{2.2cm}|C{2cm}|C{5cm}|C{5cm}|}
\caption{STRIDE-based Threat Model for Connected Vehicle Sensors} \label{tab:stride-threat-model} \\
\hline
\textbf{ID} & \textbf{Threat Description} & \textbf{STRIDE Category} & \textbf{Attack Method} & \textbf{Mitigation Recommendations} \\
\hline
\endfirsthead

\multicolumn{5}{c}{\tablename\ \thetable\ -- \textit{Continued from previous page}} \\
\hline
\textbf{ID} & \textbf{Threat Description} & \textbf{STRIDE Category} & \textbf{Attack Method} & \textbf{Mitigation Recommendations} \\
\hline
\endhead

\hline \multicolumn{5}{r}{\textit{Continued on next page}} \\
\endfoot

\hline
\endlastfoot

1 & Spoofing a sensors identity & Spoofing & An attacker can spoof one or all of the sensors' identities and provide false inputs to the controlling ECU, resulting in incorrect vehicle maneuvering actions. & All internal communication within the vehicle should be authenticated to prevent identity spoofing and the injection of malicious messages. \\ \hline
2 & Spoofing the Autonomous driving ECU identity & Spoofing & An attacker can spoof the identity of the ECU to send commands to the actuators and maneuver the vehicle. & Use authentication methods for messages transmitted over the in-vehicle communication network. \\ \hline
3 & Tampering with the message content & Tampering & An attacker listening to the in-vehicle communication network can alter message data, causing the controlling ECU to make false maneuvering decisions, which can be deadly. & All messages should include a checksum, which is then signed with a secret key to detect any tampering during transmission to their destination. \\ \hline
4 & Stealing sensor information & Information disclosure & An attacker listening to the in-vehicle network communication can intercept all data transmitted from the sensors to the controlling ECU. This data may include images from the camera sensor and other private information that should not be disclosed. & These messages should be encrypted before being transmitted over the in-vehicle communication network. This way, an attacker passively listening to the network cannot read them. \\ \hline
5 & Flooding the inner vehicle network & Denial of service & An attacker can flood the in-vehicle network, causing a denial-of-service (DoS) attack and preventing the sensors from providing any information to the controlling ECU. & A rate-limiting measure should be applied to each element in the in-vehicle network according to its expected transmission rate. \\ \hline
6 & Destroying the sensors & Denial of service & An attacker can use physical means to disrupt the sensors and make them inoperable, such as smashing a camera with a hammer or attaching a blocking element to the LIDAR sensor. Although the risk score is relatively high due to the ease of execution, we assume that in the real world it won't occur frequently. & There aren’t any real mitigations for this type of attack other than keeping your car in a secure environment. On the other hand, these attacks are loud and obvious, so we believe the likelihood of them occurring is very low. They can also be quickly detected and resolved. \\ \hline
7 & Tampering with the enviroment & Tampering & An attacker can use physical means to trick the sensors into misperceiving or misanalyzing the environment, using methods such as 2D image printing attacks, mmWave interference, and more. & The best mitigation is to use multiple sensors for decision-making, so that if one sensor is under attack, the others can detect the anomaly. \\ \hline
8 & DDOS Modem & Tampering & An attacker could impair the autonomous vehicle's modem and communication systems which would cause a disruption in safety functions including real time data. & Content Delivery Networks which will reduce the attacks on one car \\ \hline
9 & Overload LiDAR & Denial of service & The goal here to overload the LiDAR sensor to force the AV car to stop due to not being able to scan its surrounding correctly & Detection systemsFirewalls \\ \hline
10 & Prior Knowledge & Information disclosure & The attacker has prior knowledge of the modem which causes data from the av car very vulnerable & Proper authenticationRegular Security Audits \\ \hline
11 & Machine learning tampering & Tampering & An attack using machine learning could alter data which could lead to safety hazards and privacy beaches & Monitor changes in data \\ \hline
12 & Smartphone spoofing & Tampering & A smartphone could cause damage to the decision making of the autonomous vehicle when spoofing is successful. The way it would work is that spoofing fake objects in the cars optical view causing it to change its mind & GPS Signal Authentication \\ \hline
13 & Spoofing Snow & Tampering & Snow can make the cameras visually impaired by covering their lens and thus make spoofing more dangerous by hindering their accuracy. This leads to inaccuracies in visual input. & Heated Sensors \\ \hline
14 & Spoofing Rain droplets & Tampering & Rain can make the cameras visually impaired and thus make spoofing more dangerous by hindering their accuracy. & Waterproofing lens \\ \hline
15 & Fog spoofing & Tampering & The reduced visibility by fog allows the attacker to spoof fake obstacles causing the camera to mistake it for something else & AI models trained for foggy conditions \\ \hline
16 & Tampered sensors data & Tampering & The data being send over from the sensors to the software could be tampered into making the AV to make risky decisions on the software side of the car & Data Validation \\ \hline
17 & Buffer overflow attack on the Modem & Tampering & The message parsing process handles incoming messages from the outside world. An attacker can craft a specially formatted message with a malicious payload to exploit a buffer overflow vulnerability in this process and gain control over the modem. From there, the attacker can launch various attacks, as he now controls the main communication point between the vehicle and the external environment. & ollow secure coding practices for the modem, conduct code reviews, and perform penetration testing to identify vulnerabilities in the code before production. \\ \hline
18 & Software Denial of Service & Denial of service & An attacker could do a denial of service attack and overwhelm the system and exploiting vulnerabilities. The could lead the software sending commands to the ECU to potentially cause safety risks, loss of control, and put the vehicles ability to avoid collision at risk & Regular Security AuditEmployment TrainingResponse Plan \\ \hline
19 & GPS Spoofing & Spoofing & Attacker sends fake GPS signals to mislead vehicle positioning. & Use encrypted and authenticated GPS signals, sensor fusion with inertial navigation systems. \\ \hline
20 & LiDAR Sensor Blinding & Denial of Service & Strong lights or lasers blind LiDAR sensors, disrupting perception. & Use sensor redundancy, detect anomalous sensor behavior. \\ \hline
21 & Camera Image Tampering & Tampering & Physical or digital interference distorts camera inputs. & Use tamper-evident camera housings, AI-based anomaly detection. \\ \hline
22 & Radar Jamming & Denial of Service & Jamming devices block or distort radar signals. & Use frequency hopping and redundant radar systems. \\ \hline
23 & Sensor Fusion Conflict & Information Disclosure & Mismatched sensor data leads to confusion in perception system. & Cross-validate sensor inputs; trust scores for sensors. \\ \hline
24 & Adversarial Visual Input & Tampering & Special patterns confuse AI object detectors (e.g., stop sign as speed sign). & Harden perception models against adversarial attacks. \\ \hline
25 & Remote Code Injection via OTA & Elevation of Privilege & Attackers inject malicious code through software updates. & Authenticate all OTA updates, verify firmware signatures. \\ \hline
26 & Man-in-the-Middle (MITM) on V2X & Spoofing & Interception and alteration of V2X messages between vehicles. & Encrypt and authenticate all V2X communications. \\ \hline
27 & Denial of Service on CAN Bus & Denial of Service & Flooding CAN network disrupts vehicle internal communication. & Implement CAN bus segmentation and intrusion detection. \\ \hline
28 & Decision AI Model Poisoning & Tampering & Compromised training data alters decision-making algorithms. & Secure training pipelines, validate input data. \\ \hline
29 & Path Planning Misguidance & Tampering & Altering map or obstacle data to mislead vehicle path planning. & Validate route data sources; use redundant mapping systems. \\ \hline
30 & Braking System Hijack & Elevation of Privilege & Unauthorized commands issued to brake actuators. & Secure control interfaces with authentication and encryption. \\ \hline
31 & Steering Manipulation & Elevation of Privilege & External commands take over steering functions. & Isolate critical control systems; use hardware security modules. \\ \hline
32 & Fake V2V Message & Spoofing & Fake messages mislead the vehicle about nearby traffic. & Use cryptographic authentication of V2V messages. \\ \hline
33 & Cloud Data Breach & Information Disclosure & Unauthorized access to AV logs or personal data stored in cloud. & Encrypt data at rest and in transit; enforce strict access control. \\ \hline
34 & Insecure API Exposure & Tampering & Open APIs allow unauthorized commands or data extraction. & Use API gateways with authentication and rate limiting. \\ \hline
35 & Unencrypted V2X Transmission & Information Disclosure & Sensitive V2X data transmitted without encryption can be intercepted. & Encrypt all V2X communications using secure protocols. \\ \hline
36 & Roadside Unit Compromise & Tampering & Manipulating traffic infrastructure to send false data to AVs. & Physically secure RSUs, use mutual authentication protocols. \\ \hline
37 & Physical ECU Tampering & Elevation of Privilege & Accessing and reprogramming critical vehicle ECUs. & Use tamper-proof enclosures, monitor for physical breaches. \\ \hline
38 & Insider Threat in Manufacturing & Elevation of Privilege & Malicious insiders embedding vulnerabilities in hardware/software. & Enforce strict supply chain audits and access controls. \\ \hline
39 & Voice Command Exploit & Spoofing & Exploiting voice interfaces to issue unauthorized commands. & Use voiceprint authentication; limit sensitive voice actions. \\ \hline
40 & Cloud-to-AV Sync Attack & Tampering & Delaying or tampering with data syncing between cloud and AV. & Authenticate sync messages; use integrity checks. \\ \hline
41 & Privacy Violation via Telemetry Data & Information Disclosure & Personal location or behavior data leaked through telemetry. & Anonymize telemetry data; comply with data protection laws. \\ \hline
42 & False Firmware Update Notification & Spoofing & Fake update prompts trick users into installing malware. & Use secure notification mechanisms; validate updates cryptographically. \\ \hline
\end{longtable}

\clearpage
\onecolumn

\clearpage
\onecolumn

\begin{figure}[!htbp]
    \centering
    \includegraphics[width=0.95\textwidth]{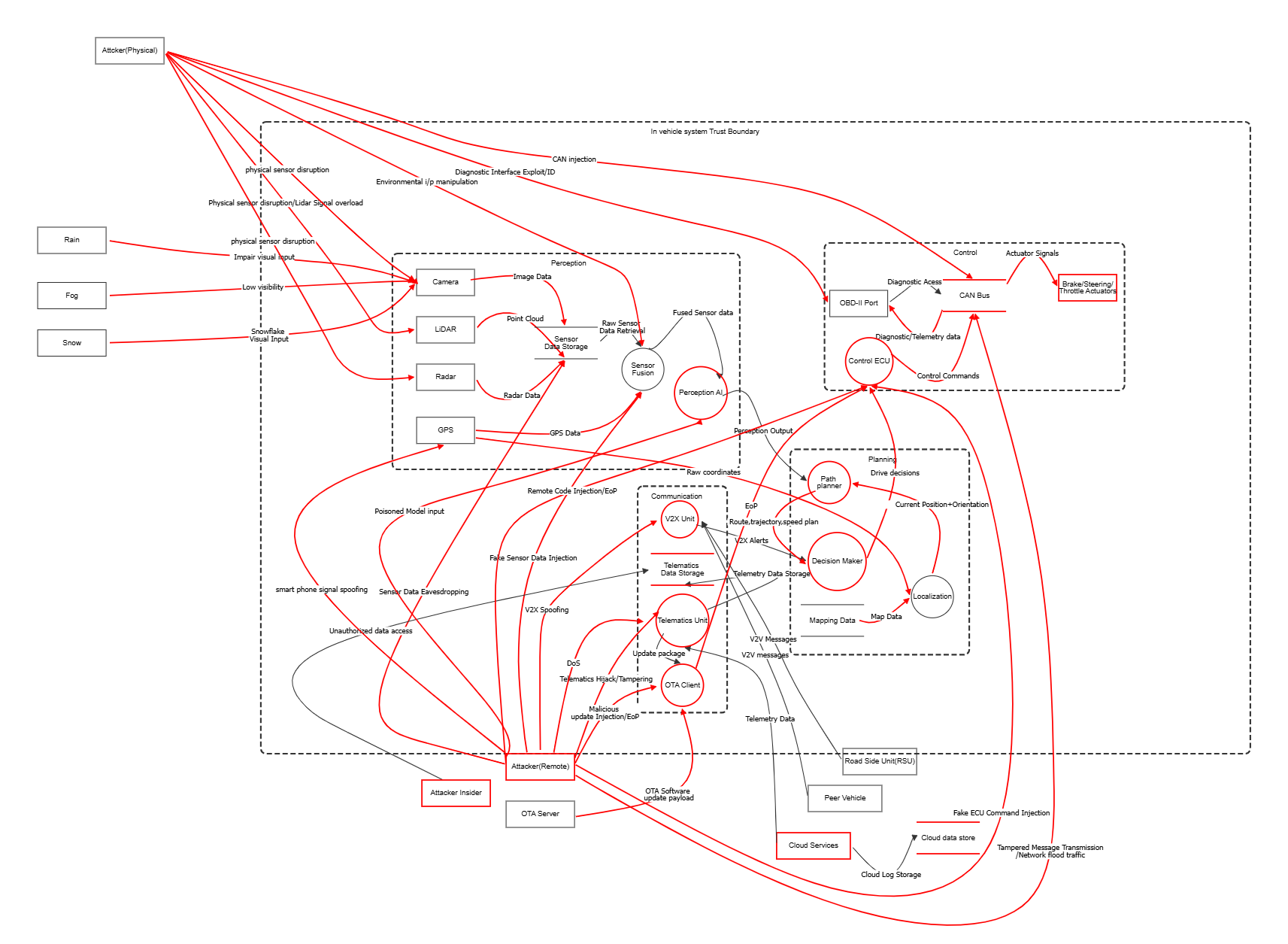}
    \caption{Autonomous Vehicles Threat Model}
    \label{fig:threat-model}
\end{figure}

\twocolumn

\twocolumn
\section{Explanation}
The threat model developed for the autonomous vehicle (AV) system provides a comprehensive and structured analysis of the various cybersecurity risks inherent to the AV ecosystem. The model incorporates external actors, internal processes, and critical data stores to create a complete representation of the system's operational and security landscape.

\textbf{External Actors and Threat Origins}

External actors modeled in the system include the Remote Attacker, Physical Attacker, Insider Attacker, Natural Environment (including Rain, Fog, and Snow), Over-the-Air (OTA) Server, Roadside Units (RSUs), Peer Vehicles, and Cloud Services. These actors were strategically selected to represent diverse threat origins that operate outside the trusted vehicle boundary. Each of these entities has the potential to inject malicious data, exploit system vulnerabilities, or disrupt communication protocols. For instance, Remote Attackers may attempt code injection through OTA updates, while natural environmental conditions can impair sensor performance.

\textbf{Internal Processes and Functional Modules}

The internal architecture of the AV system includes essential processes such as the Sensor Fusion Unit, Perception AI, V2X Communication Unit, Telematics Unit, OTA Client, Path Planner, Decision Maker, Localization Module, Mapping Data Process, and the Control ECU. These processes collectively enable the AV to perceive its environment, make real-time decisions, and interact with both internal subsystems and external entities. Threats against these components were assessed based on their role in ensuring the safety, reliability, and autonomy of the vehicle.

\textbf{Critical Data Stores and Data-at-Rest Risks}

Key data stores include Sensor Data Storage, Telematics Data Storage, V2X Data Buffer, Perception Output Storage, Mapping Data Storage, OTA Update Storage, and Control Commands Log. These repositories were modeled to reflect the potential risks associated with data-at-rest, such as unauthorized access, data tampering, or leakage of sensitive operational information. The presence of these data stores is vital for maintaining system continuity, traceability, and diagnostics.

\textbf{Threat Identification and Classification Using STRIDE}

Each threat was mapped to a corresponding system element using the STRIDE classification framework, which categorizes threats as Spoofing, Tampering, Repudiation, Information Disclosure, Denial of Service (DoS), and Elevation of Privilege. This structured approach enabled systematic identification and documentation of vulnerabilities across the AV system.

\textbf{
Risk Prioritization Through DREAD Scoring}

The threats were evaluated and prioritized using the DREAD risk assessment model, which measures the Damage potential, Reproducibility, Exploitability, Affected users, and Discoverability of each threat. This provided a quantifiable risk score to rank threats based on their impact and likelihood. High-severity threats such as GPS Spoofing, CAN Bus Denial of Service, Cloud Data Breach, Remote Code Injection via OTA, and Decision AI Model Poisoning were identified as critical due to their ability to compromise vehicle safety, functionality, and user trust.

\textbf{Mitigation Strategies and Controls}

Tailored mitigation strategies were developed for each threat category. Spoofing threats were countered with mutual authentication protocols, digital certificates, and encrypted communications. Tampering threats were addressed through anomaly detection algorithms, redundant sensing mechanisms, and secure software engineering practices. DoS threats were mitigated via network segmentation, rate-limiting, and the deployment of Intrusion Detection Systems (IDS). Information Disclosure risks were mitigated by enforcing encryption standards, strict access control policies, and secure cloud integration. Elevation of Privilege threats were mitigated through signed firmware validation, hardware-level security modules, and role-based access control mechanisms.

\textbf{
Modeling of Environmental Conditions as Threats}

Environmental conditions such as Rain, Fog, and Snow were modeled as external actors with potential to degrade perception accuracy. These threats, categorized under Tampering, affect the fidelity of sensor inputs and thereby impair situational awareness. Although these threats are not malicious in nature, they pose significant safety concerns and were rated as medium in severity under DREAD. Mitigations included hardware-based solutions (heated and waterproofed sensors), software resilience (AI models trained for adverse conditions), and system-level redundancy.

\textbf{Final Threat Model Overview}

The final threat model incorporates a total of forty-two distinct threats, each precisely mapped to system components and prioritized based on severity. This comprehensive coverage enabled focused risk mitigation planning and design reinforcement, particularly for top-ranked threats such as Adversarial Visual Input, GPS Spoofing, Cloud Data Breach, CAN Bus Denial of Service, and Decision AI Model Poisoning. These threats were addressed with heightened attention due to their potential to undermine safety-critical functions of the AV system.

\textbf{
Standards Alignment and Cybersecurity Best Practices}

The selected mitigations were aligned with recognized cybersecurity frameworks and industry standards, including AUTOSAR security guidelines, the OWASP Top Ten for Connected Vehicles, and ISO/SAE 21434 standards for automotive cybersecurity. This ensured that the proposed threat responses were both technically robust and compliant with regulatory expectations.

\section{Conclusion}

Through this threat modeling process, the AV system was evaluated from a holistic cybersecurity perspective. The model provides a foundation for secure system design, proactive defense planning, and continuous risk monitoring in the face of both cyber and physical threat vectors. By embedding security into the system architecture, the AV platform is better prepared for safe and reliable deployment in real-world environments.

\vspace{12pt}


\section{References}

\begin{thebibliography}{20}

\bibitem{ref1}
Hoque, T., Hasan, R., \& Aminul, M. (n.d.). \textit{Autonomous driving security: A comprehensive threat model of attacks and mitigation strategies}. Retrieved from \url{https://par.nsf.gov/servlets/purl/10399046}

\bibitem{ref2}
Zikria, P. G., Lyu, Z. P., \& Choi, H. L. (2019). Autonomous vehicle security: Conceptual model. \textit{IEEE Access, 7}, 92356–92367. \doi{10.1109/ACCESS.2019.2924889}

\bibitem{ref3}
Chakraborty, T., \& Chen, Q. A. (2024). \textit{Risk assessment for security threats and vulnerabilities of autonomous vehicles (UC-ITS-RIMI-5B-03)}. Institute of Transportation Studies. \doi{10.7922/G2N29V87}

\bibitem{ref4}
Zhang, L., Li, X., \& Zhou, Q. (2021). Digital twin analysis to promote safety and security in autonomous vehicles. \textit{IEEE Transactions on Industrial Informatics, 17}(5), 2355–2363. \doi{10.1109/TII.2021.3078901}

\bibitem{ref5}
Day, T. S. (2022, November 22). \textit{Spoofing LiDAR could blind autonomous vehicles to obstacles}. Retrieved from \url{https://hackaday.com/2022/11/22/spoofing-lidarcould-blind-autonomous-vehicles-to-obstacles/}

\bibitem{ref6}
Smith, J. P., Williams, A. S., \& Patel, K. R. (2021). A systematic risk assessment framework of automotive cybersecurity. \textit{Journal of Automotive Security, 3}(2), 153–164.

\bibitem{ref7}
McLellan, D. P., Seo, D. H., \& Zhang, S. B. (2022). Internet of Autonomous Vehicles Communications Security: Overview, Issues, and Directions. \textit{IEEE Wireless Communications, 26}(4), 60–65. \doi{10.1109/MWC.2019.1800503}

\bibitem{ref8}
Zikria, P. G., Choi, H. L., \& Al-Surmi, M. J. (2021). Analysis of automotive security risk using cyber security. \textit{IEEE Access, 9}, 33145–33156. \doi{10.1109/ACCESS.2021.3064567}

\bibitem{ref9}
Wang, F. M., Lee, J. B., \& Zheng, X. F. (2022). PIER: Cyber-resilient risk assessment model for connected and autonomous vehicles. \textit{Journal of Cyber-Physical Systems, 7}(4), 451–463. \doi{10.1007/s11276-022-03084-9}

\bibitem{ref10}
Liu, M. F., Fisher, J. R., \& Singh, W. L. (2021). A comprehensive threat model for autonomous vehicle communication security. Unpublished manuscript. Retrieved from \url{https://www.researchgate.net/publication/abcdef}

\bibitem{ref11}
Pandya, R. (2021). Cybersecurity engineering: Bridging the security gaps in advanced automotive systems and ISO/SAE 21434. \textit{IEEE Transactions on Industrial Informatics, 17}(5), 3456–3464. \doi{10.1109/TII.2021.3067894}

\bibitem{ref12}
STRIDE threat model-based framework for assessing the vulnerabilities of modern vehicles summary. (2023, October). \textit{Computer Security, 133}. Retrieved from \url{https://doi.org/10.1016/j.cose.2023.102456}

\bibitem{ref13}
Nanda, A., Puthal, D., Rodrigues, J. J. P. C., \& Kozlov, S. A. (2019). Internet of Autonomous Vehicles Communications Security: Overview, Issues, and Directions. \textit{IEEE Wireless Communications, 26}(4), 60–65. \doi{10.1109/MWC.2019.1800503}

\bibitem{ref14}
Novelli, C., Casolari, F., Rotolo, A., Taddeo, M., \& Floridi, L. (2023). Taking AI risks seriously: A new assessment model for the AI Act. \textit{AI \& Society}. \doi{10.1007/s00146-023-01723-z}

\bibitem{ref15}
Sun, Z., Su, L., Balakrishnan, S., Bhuyan, A., Wong, P., \& Qiao, C. (2021). [Title unavailable]. \textit{IEEE Xplore}. Retrieved from \url{https://ieeexplore.ieee.org/Xplore/home.jsp}

\bibitem{ref16}
Wang, Y., Wang, Y., Qin, H., Ji, H., Zhang, Y., \& Wang, J. (2021). A systematic risk assessment framework of automotive cybersecurity. \textit{Automotive Innovation}. \doi{10.1007/s42154-021-00140-6}

\bibitem{ref17}
Zhang, Y. Y., Luo, X. X., \& Lin, Z. Z. (2020). Bird’s-eye view on the automotive cybersecurity landscape: Challenges in adopting AI/ML. \textit{IEEE Access, 8}, 101080–101089. \doi{10.1109/ACCESS.2020.2998947}

\bibitem{ref18}
Lee, F. P., Zheng, H. F., \& Fang, C. S. (2021). STRIDE: Scalable and secure over-the-air software update scheme for autonomous vehicles. \textit{IEEE Access, 9}, 105789–105798. \doi{10.1109/ACCESS.2021.3056789}

\bibitem{ref19}
An Integrated Approach of Threat Analysis for Autonomous Vehicles Perception System. (2023). \textit{IEEE Journals Magazine}. Retrieved from \url{https://ieeexplore.ieee.org/document/10041909/}

\bibitem{ref20}
Zhang, Z. S., Huo, Y. L., \& Wang, L. (2021). Who is in control? Practical physical layer attack and defense for mmWave-based sensing in autonomous vehicles. \textit{IEEE Transactions on Vehicular Technology, 70}(6), 4574–4583. \doi{10.1109/TVT.2021.3061234}

\bibitem{ref21}
Dhakal, S., Qu, D., Carrillo, D., Yang, Q., \& Fu, S. (2021). OASD: An open approach to self-driving vehicle. \textit{2021 Fourth International Conference on Connected and Autonomous Driving (MetroCAD)}, Detroit, MI, USA, 54–61. \doi{10.1109/MetroCAD51599.2021.00017}

\bibitem{ref22}
Tezerjani, M. D., Qu, D., Dhakal, S., Carrillo, D., Mirzaeinia, A., \& Yang, Q. (2025). Real-time motion planning for autonomous vehicles in dynamic environments. In D. D. Hodson, M. R. Grimaila, H. R. Arabnia, L. Deligiannidis, \& T. J. Wagner (Eds.), \textit{Scientific computing and bioinformatics and computational biology. CSCE 2024. Communications in computer and information science} (Vol. 2258). Springer, Cham. \doi{10.1007/978-3-031-85902-1\_14}


\end{thebibliography}
\end{document}